\documentclass[sigconf]{acmart}

\AtBeginDocument{%
  \providecommand\BibTeX{{%
    \normalfont B\kern-0.5em{\scshape i\kern-0.25em b}\kern-0.8em\TeX}}}

\copyrightyear{2025}
\acmYear{2025}
\setcopyright{cc}
\setcctype{by}
\acmConference[C\&C '25]{Creativity and Cognition}{June 23--25, 2025}{Virtual,
United Kingdom}
\acmBooktitle{Creativity and Cognition (C\&C '25), June 23--25, 2025, Virtual,
United Kingdom}\acmDOI{10.1145/3698061.3726913}
\acmISBN{979-8-4007-1289-0/2025/06}

\usepackage{array, multirow}
\usepackage{enumitem}  
\usepackage{tabularx}
\usepackage{hyperref}
\usepackage{etoolbox}
\begin{document}

\title{Coping with Uncertainty in UX Design Practice: Practitioner Strategies and Judgment}

 
\author{Prakash Chandra Shukla}
\email{shukla37@purdue.edu}
\affiliation{%
 \institution{Purdue University}
 \city{West Lafayette}
 \state{Indiana}
 \country{USA}
}
\author{Phuong Bui}
\email{bui42@purdue.edu}
\affiliation{%
 \institution{Purdue University}
 \city{West Lafayette}
 \state{Indiana}
 \country{USA}
}
\author{Paul Parsons}
\email{parsonsp@purdue.edu}
\affiliation{%
 \institution{Purdue University}
 \city{West Lafayette}
 \state{Indiana}
 \country{USA}
}

\renewcommand{\shortauthors}{Shukla, et al.}

\begin{abstract}
  The complexity of UX design practice extends beyond ill-structured design problems to include uncertainties shaped by shifting stakeholder priorities, team dynamics, limited resources, and implementation constraints. While prior research in related fields has addressed uncertainty in design more broadly, the specific character of uncertainty in UX practice remains underexplored. This study examines how UX practitioners experience and respond to uncertainty in real-world projects, drawing on a multi-week diary study and follow-up interviews with ten designers. We identify a range of practitioner strategies---including adaptive framing, negotiation, and judgment---that allow designers to move forward amid ambiguity. Our findings highlight the central role of design judgment in navigating uncertainty, including emergent forms such as temporal and sacrificial judgment, and extend prior understandings by showing how UX practitioners engage uncertainty as a persistent, situated feature of practice.
\end{abstract}





\maketitle

\section{Introduction}

Uncertainty is common across many professional fields, shaping the complex and dynamic environments in which practitioners operate \cite{griffin_when_2020}. In design, it is both a persistent challenge and a potential catalyst for innovation. Daalhuizen et al. \cite{daalhuizen_dealing_2009} argue that contemporary design research must better equip practitioners to navigate uncertainty in practice. Scholars from diverse disciplines, including management research \cite{sniazhko_uncertainty_2019, frishammar_beyond_2011}, policy-making \cite{walker_dealing_2003}, medicine \cite{oriordan_dealing_2011}, engineering \cite{pelz_mastering_2021}, industrial practice \cite{rocquigny_uncertainty_2008}, and decision sciences \cite{ove_hansson_decision_1996}, have explored how uncertainty influences professional practice. Yet, despite its centrality, design lacks a standardized approach to managing uncertainty or fostering innovation. Stolterman \cite{stolterman_challenge_2021} highlights the inherent challenges of attempting to control  design processes, urging researchers to address this complexity more directly. This positions the study of uncertainty as foundational to advancing design research and practice.

Design research has long grappled with the uncertainties inherent in defining and addressing design problems. A central concern has been the indeterminate nature of design situations and the challenges they pose for practitioners \cite{simon_structure_1973, rittel_dilemmas_1973, suchman_plans_1987, buchanan_wicked_1992, dorst_creativity_2001, stolterman_nature_2008}. Simon \cite{simon_structure_1973} described design problems as \textit{`ill-structured'}, requiring \textit{`satisficing'} rather than optimizing. Schön \cite{schon_reflective_1983} instead argued that design problems are not definable from the outset, underscoring the designer's role in framing the problem and managing the inherent complexity. Rittel and Webber \cite{rittel_dilemmas_1973} characterized design problems as `wicked' due to their complexity and resistance to straightforward solutions. Building on this, Dorst \cite{dorst_exploring_2003} suggested that understanding design behavior in practice requires acknowledging the situated nature of design problems, and further emphasized \cite{dorst_problem_2004} the importance of viewing these problems through the eyes of the designer. From this perspective, problem and solution spaces often co-evolve \cite{dorst_creativity_2001}, meaning that uncertainty is not an anomaly but a defining condition of design practice.

While uncertainty is intrinsic to design problems \cite{rittel_wicked_1967}, the realities of contemporary design practice introduce additional and often compounding forms of uncertainty. UX practitioners commonly juggle multiple projects; coordinate with various stakeholders; advocate their own value within organizations; and work with new and changing tools, all in resource-constrained environments \cite{stolterman_nature_2008, tracey_uncertainty_2016, jones_determinants_2019, crescenzi_impacts_2016}. These practical demands generate new pressures not fully captured by traditional theories of design cognition and practice. For instance, Goodman et al. \cite{goodman_understanding_2011} describe how practitioners face uncertainty in managing and appeasing clients. Norouzi et al. \cite{norouzi_architect_2015} discuss communication challenges and the resulting uncertainty in client-architect relationships. Fedosov et al. \cite{fedosov_challenges_2021} articulate the challenges in transferring UX designs across products due to coordination difficulties among diverse teams with conflicting goals. This work points to the entangled, collaborative, and often unstable contexts in which UX work is embedded---contexts that shape how uncertainty is produced, perceived, and negotiated in practice.

While researchers have examined some factors that contribute to uncertainty faced by practitioners across design fields, there remains a lack of inquiry into how uncertainty manifests in UX practice specifically. This gap is increasingly consequential given recent shifts in UX design practice, including evolving roles, expanded responsibilities, and the changing nature of what constitutes UX knowledge \cite{branch_user_2021, kou_practice-led_2019, macdonald_how_2022}. Researchers in fields such as product design \cite{de_weck_classification_2007}, project management \cite{saunders_conceptualising_2015} and decision sciences \cite{lipshitz_coping_1997} have developed typologies and structured frameworks to account for uncertainty. These efforts have helped clarify the conditions under which uncertainty arises and how it can be productively managed. Yet in UX, no equivalent conceptual scaffolding exists, despite the profession’s growing complexity and increasing entanglement with legal, organizational, and technological constraints. As UX continues to evolve, there is a pressing need to understand not just that uncertainty exists, but how practitioners perceive, interpret, and respond to it in practice. 


In this research, we investigated how UX designers experience and respond to uncertainty in their everyday practice. We employed a four-week-long diary study with 10 UX professionals, followed by semi-structured interviews that explored the contextual dynamics, decision points, and perceived stakes surrounding moments of uncertainty. Our aim is not only to document the types of uncertainty that arise, but to understand how practitioners interpret and work through these conditions in situ. This study is guided by the following research questions: \begin{itemize}
    \item What uncertainties do UX designers encounter in their everyday practice?  
    \item What strategies do UX practitioners employ to handle uncertain situations?
\end{itemize}

We contribute a practice-grounded account of how uncertainty is encountered and managed. We extend prior design research by characterizing the forms of judgment, framing, and negotiation that practitioners rely on---not simply as responses to disruption, but as constitutive elements of design work under real-world conditions.

\section{Background}
\subsection{Design as Situated Practice under Uncertainty} 
At its core, design represents a practice of \textit{`intentional change'} \cite{nelson_design_2012}, where practitioners aim to create new realities that do not yet exist. Unlike traditional scientific methods, design processes are emergent and contingent, unfolding in response to evolving constraints, goals, and interpretations \cite{stolterman_nature_2008, schon_problems_1984, cross_design_1981}. Designers often operate in what Schön termed \textit{``messy situations''} \cite{schon_reflective_1983}, which are typically complex and under-determined with potentially infinite sources of information, constraints, and possibilities \cite{rittel_dilemmas_1973, buchanan_wicked_1992}. In such settings, design is not simply problem-solving; rather, it involves framing what is at stake and how to proceed \cite{schon_reflective_1983}. This framing is especially consequential in UX, where success is not measured by adherence to a process, but by the subjective and sometimes contested experiences a product elicits \cite{stolterman_nature_2008, mccarthy_technology_2007}.

In practice, designers work with demanding clients, limited information, new technologies, tight deadlines, and insufficient tools, adding layers of uncertainty to their work \cite{stolterman_nature_2008}. Gray \cite{gray_its_2016} notes that practitioners often adapt and modify standard design methods to suit evolving needs, revealing a mode of practice distinct from the procedures often emphasized in academic accounts. 

\subsubsection{Evolution of UX Design Practice}
In recent years, the role of UX design practitioners has evolved, encompassing management tasks beyond traditional design responsibilities and expanding the scope of demands and uncertainties they face \cite{meyer_changing_2020, gray_legal_2024, takaffoli_generative_2024}. This evolution is driven by the increasing pervasiveness of technology in daily life and the growing recognition of the user's experience as critical to product success \cite{cajander_ux_2022, kou_practice-led_2019}. UX practitioners are increasingly expected to navigate multiple emergent and dynamic topics, like AI/ML \cite{takaffoli_generative_2024, yang_investigating_2018, zdanowska_study_2022}, regulatory and legal frameworks \cite{gray_legal_2024}, dark patterns and other ethical concerns \cite{gray_dark_2023, sanchez_chamorro_ethical_2023}, and tensions between their design philosophy and practical realities \cite{watkins_tensions_2020}. While the professional identity of UX design has achieved some stability in recent years, the demands and uncertainties of everyday practice are still in flux. Prior accounts from related design domains, including architecture, engineering design, and interaction design, do not fully capture the situated complexities of contemporary UX practice. Understanding how practitioners interpret and respond to uncertainty in these conditions is critical for both research and education. 

\subsection{Disciplinary Perspectives on Uncertainty}
Uncertainty has long been recognized as a defining feature of professional practice, describing situations where knowledge is insufficient to predict outcomes  \cite{saunders_conceptualising_2015, bar-anan_feeling_2009}. Across disciplines, scholars have proposed models to classify and manage uncertainty, distinguishing between informational gaps, ambiguity in goals, and instability in conditions \cite{lipshitz_coping_1997, mikkelsen_perceived_2021, lipshitz_coping_1997}.


In design, uncertainty is constitutive. Researchers have emphasized the role of framing, abductive reasoning, and co-evolution in navigating design's indeterminate nature  \cite{dorst_core_2011, dorst_frame_2015, stolterman_nature_2008}. Soden et al. \cite{soden_modes_2022} describe four relational modes of uncertainty in HCI: taming, generative, political, and lived experience. These perspectives move beyond treating uncertainty as a problem to solve, instead framing it as a condition to engage through practice. While some studies have noted that UX practitioners adapt methods due to constraints like limited user access or tight timelines \cite{cajander_ux_2022}, we still lack a grounded account of how uncertainty is perceived and navigated in UX practice. As the scope and stakes of UX work expand, addressing this gap is important---not only for research, but for cultivating the practical judgment required in design education and professional development. 

\section{Methods} 
This study investigates how UX designers experience and navigate uncertainty in their everyday work. We adopted an interpretivist, practice-oriented approach grounded in qualitative inquiry. Our aim was not to produce a comprehensive typology of uncertainty across the entire UX profession, but to generate insight into how experienced practitioners interpret, respond to, and work through uncertainty in context. To that end, we combined a four-week diary study with follow-up semi-structured interviews, enabling us to elicit situated accounts of uncertainty and examine how designers reflect on their own strategies and judgments in response.

\subsection{Participants} We recruited 10 UX designers globally through a screening survey distributed on LinkedIn, Twitter, and UX-related subreddits. Participants provided a design portfolio or LinkedIn profile for verification. From over 200 responses, 20 invitations were sent, selecting only practicing UX professionals fluent in English. Individual contributors actively engaged in UX design were included, while those in purely managerial roles were excluded. We reviewed each participant's profile to verify authenticity. While UX practitioners hold various titles \cite{vinney_11_2022}, all participants self-identified as UX designers. Participants varied in role title, level of experience, and organizational context. They held positions such as Product Designer, UX Designer, UI/UX Designer, and Design Consultant, with experience levels ranging from 2 to 11+ years. All were currently working in industry, such as finance, healthcare, e-commerce, civic tech, enterprise software, and non-profit services. While this group does not represent the full diversity of UX contexts, it reflects the environments in which UX practitioners commonly encounter both strategic and execution-level uncertainties. Of 20 invited, 14 agreed but four dropped out during the diary study. See Table \ref{tab:demographics} for details.

\subsection{Data Collection}
Participants completed weekly diary entries over four weeks using an online form. Each week, they described a situation involving uncertainty—cases where the next step was unclear, stakes were ambiguous, and no clear solution was available. Prompts asked about the nature of the uncertainty, their thought process, actions taken, and outcomes. These entries served as generative probes that sensitized participants to the role of uncertainty and provided material for deeper reflection.

Following the diary phase, we conducted semi-structured Zoom interviews (35–55 minutes). Interviews began by revisiting selected diary entries and then explored broader patterns and reflections. Our protocol was informed by the Critical Decision Method (CDM) \cite{klein_critical_1989}, which we adapted for UX contexts to elicit narratives of non-routine, interpretive decision-making. We explored how participants perceived uncertainty, made sense of ambiguous situations, and exercised judgment over time.


\begin{table}
  \caption{Demographics of our study participants.}
  \label{tab:demographics}
  \begin{tabular}{cccc}
    \toprule
    Participant & Title & YoE & Region\\
    \midrule
    P1 & Freelance Designer & 11+ & Americas \\
    P2 & Sr. UX Designer & 6 - 10 & Europe \\
    P3 & Product Designer & 2 - 5 & Asia-Pacific \\
    P4 & Product Designer & 2 - 5 & Europe \\
    P5 & Design Leader & 11+ & Americas \\
    P6 & Design Consultant & 11+ & Asia-Pacific \\
    P7 & Sr. Product Designer & 6-10 & Europe\\
    P8 & Sr. UX Designer & 11+ & Asia-Pacific \\
    P9 & Sr. Product Designer & 6-10 & Asia-Pacific \\
    P10 & UX/UI Designer & 2 - 5 & Asia-Pacific \\
  \bottomrule
  \multicolumn{4}{c}{\footnotesize Note: YoE - Years of Experience; Sr. - Senior}
\end{tabular}
\end{table}

\subsection{Data Analysis} We analyzed all diary entries and interview transcripts using a hybrid thematic approach \cite{fereday_demonstrating_2006}, combining inductive and deductive coding. Deductive codes drew on prior work in design complexity and judgment (e.g., \cite{stolterman_nature_2008, schon_reflective_1983, garvey_design_2016, rittel_dilemmas_1973}). This resulted in the identification of the following codes: stakeholder management, time constraints, resource constraints, and wicked problems. These codes provided a structured framework for an initial round of analysis. 

We adopted Swain's three-phase model for hybrid thematic analysis \cite{swain_hybrid_2018}. All authors contributed to the analysis, with the third author assuming a supervisory role. The first phase involved familiarizing ourselves with the data by thoroughly reviewing the diary entries and interview transcripts. In the second phase, we applied a priori codes while conducting inductive analysis to generate a posteriori codes. Two authors independently coded transcripts and diary entries, then cross-checked and resolved discrepancies. Regular team meetings ensured consistency. Inductive coding was data-driven yet informed by design theory. The process was managed using Dovetail for collaborative analysis. By the second phase’s conclusion, we identified two key themes: ``Types of Uncertainty'' with 10 subthemes and ``Methods to Tackle Uncertainty'' with 8 subthemes. In the third phase, these themes were refined, consolidating overlapping subthemes through iterative discussions. This process ensured a more focused and coherent presentation of the study’s findings.




\section{Findings}
Our analysis surfaced two overarching themes: (1) the types of uncertainty UX practitioners encounter in their work, and (2) the strategies they employ to interpret, manage, or mitigate these uncertainties. These findings respond directly to our research questions and build a practice-grounded account of how uncertainty is experienced and navigated in real-world UX contexts. 

We first describe five recurring sources of uncertainty, derived through thematic analysis across diary entries and interview transcripts. These include: people \& power dynamics, collaborative processes, operational planning, project (re)framing, and feasibility constraints. We then turn to the strategies designers use to navigate these uncertain conditions, including adaptive framing, negotiation, improvisation, and expectation-setting. These categories are not mutually exclusive; instead, they serve as analytic lenses that help surface patterns across participant accounts. In many cases, uncertainty arises at the intersection of multiple domains, such as stakeholder dynamics contributing to planning instability, or feasibility constraints shaping project reframing. Our goal is not to define rigid types, but to reveal how practitioners narrate and reason through uncertainty in complex, situated contexts. Table~\ref{tab:typology} provides an overview of the relationship between uncertainty sources and practitioner strategies.
 

\subsection{Where Uncertainty Arises in Practice}
\subsubsection{People \& Power} 
Uncertainty often stems from power asymmetries between UX practitioners and stakeholders, clients, or decision-makers. These situations are marked not simply by differing opinions, but by misalignments between authority and expertise. Stakeholders may impose expectations or directives without a clear understanding of UX processes or user needs, leaving designers to reconcile conflicting priorities or respond to demands that lack grounding. Uncertainty emerges not only from what is unknown, but from what is contested—who gets to define problems, set priorities, or validate design directions.  These are not isolated incidents but recurring patterns that reflect structural ambiguities in how UX is situated within organizational hierarchies.

P7 illustrates this with a client's sudden shift in direction: \textit{``[The client] said, This needs to be more stylish,' which was completely the opposite of what we expected. It was strange that they requested it, we implemented it, and then in the end, [the client] came back, and everything changed again.''} Similarly, P8 describes the volatility of evolving stakeholder input: \textit{``they've had different meetings with [other stakeholders] and the requirements kept changing.''} These changes often originate not from new user data but from shifting internal politics or top-down mandates. As P3 highlights: \textit{``Suddenly, the product manager insists that this is an important product to be shipped ... because it was a business need.''}

In some cases, stakeholders assert control based on positional authority rather than product expertise. P5 reflects on this challenge: \textit{``There are opinionated [clients] ... But when they take control, despite never having made a digital product before—only running the company—there’s a problem. They think they know what users want, and their solutions are often extremely painful.''} P8 describes a similar dynamic: \textit{``[The client] said, `Look, I've spent three years on this. This is what I want on the page, and I'm confident. I know this is what people want because, as a finance expert, it's what I'd want to see.'''} Others noted that stakeholder expectations are often shaped by design trends or buzzwords, with little connection to user needs.  P7 recalls a vague request: \textit{``they might say, I want gamification in our app.' When asked why or what they needed, they didn’t have a clear answer. They only know the term and think it might be useful for the product.''} 

\subsubsection{Collaboration \& Coordination}
Another source of uncertainty stems from working across teams with differing norms, misaligned expectations, or incompatible communication styles. Unlike uncertainty tied to external stakeholders, these challenges arise within the design organization itself. As designers move between teams or join ongoing projects, they must quickly adapt to unfamiliar practices and rhythms, often without formal onboarding or shared process infrastructure.

P8 describes the challenge of adapting to new team environments: \textit{``The biggest challenge for me was joining a six-week or three-month project and learning how they worked---their formats, expected deliverables, and how to communicate effectively within their team.''} P6 echoes this, noting that even understanding how a team prefers to collaborate can be uncertain: \textit{``It was a new team working together for the first time... there were a lot of uncertainties about processes and how we preferred to work together.''} These difficulties are not limited to small or ad hoc teams. P5 notes that even large, well-resourced companies suffer from misalignment: \textit{``Most design is very fast. However, for a huge company, sometimes it takes long. It could be six months to get in the groove with them.''}

Uncertainty also emerges around deliverables and approvals, especially when roles and responsibilities are diffuse. P2 highlights recurring confusion over what a wireframe represents: \textit{``[Clients] often return with design feedback on the wireframe, indicating we haven't fully explained a wireframe. Additionally, they’re unclear on the sign-off process.''} In one case, no one had formal authority to approve work: \textit{``The problem was that being a relatively young team with no expertise in the digital area, it was unclear to [the team] who exactly was responsible for approval. One week, we came up with the set of features designed from our ideation session and there was no one responsible for signing off.''} P3 recounts how this kind of ambiguity led to drawn-out revisions: \textit{``It [the homepage redesign] involved numerous meetings and revisions, and it eventually became an in-house joke: We're redoing this menu again. It's the world's most expensive menu.''}

At times, these coordination failures reflect deeper fractures in organizational coherence.  P1 recounts a striking misalignment around company values: \textit{``Two-thirds of the company said they have a general idea of the company’s values but don't know them exactly. One-third claimed they could recite the values and that these values influence their daily work. However, even within this group, no single value was unanimously agreed upon, they were completely unaligned.''} These internal disconnects contribute to a form of systemic uncertainty—one not rooted in technical complexity, but in the social infrastructure of the design process itself.

\subsubsection{Process \& Planning}
Uncertainty also arises from discontinuities in project structure, unstable timelines, and unclear planning processes. Designers may be pulled into projects midstream or asked to manage shifting priorities across multiple streams. P1 describes the disorientation of joining an in-progress project without context: \textit{``I've joined projects where the train has already left the station, and I'm just running to catch up. ... User journeys, research plans, and a whole bunch of documentation to catch up on.''} Fragmented project transitions compound these issues, as P3 highlights: \textit{``I shifted from that project to the other one. My product manager did the same. Only the engineering team continued to work on that. When we came back, the engineers said that this might not work, making a huge difference for our intended timeline.''} These discontinuities are intensified when designers are stretched across multiple initiatives. P4 describes the difficulty of managing multiple projects simultaneously: \textit{``I work with two, sometimes three product managers at the same time. Sometimes, we fail to align and I have to split my working hours for different projects ... They all become a priority because every manager wants their project feature to be launched first.''} Such fragmentation introduces uncertainty not just about delivery timelines, but about how to allocate attention and justify trade-offs.

In some cases, the absence of formal planning infrastructure further compounds uncertainty.  P1 recalls a project where a delay had no discernible consequence: \textit{``There was no timeline ... We had a list of tasks we were behind on, but there was no visualization of how those delays would impact the project or how one delay would affect other areas. There was no connection between these elements.''} Others faced compressed timelines that rendered normal processes infeasible. As P3 puts it: \textit{``Any project at this level requires at least three months for discovery research, conducting interviews, gaining insights, and actual design work ... Yet, I was given 20 days initially.''}

Resource limitations, both human and technical, add further uncertainty. P9 describes having to take on customer support responsibilities due to staffing gaps: \textit{``We don't have a customer support team. So actually me, myself, and I have to research the tool which is affordable for the company''} P7 describes how, in some cases, designers have limited user access: \textit{``in consultancy, it's hard to get into the user part. You can sometimes, but not always.''} These constraints reshape what kinds of processes are even possible, forcing designers to adapt their methods or assumptions in the face of incomplete planning infrastructure.

\subsubsection{Project (Re)framing}
A further source of uncertainty emerges from ambiguity about what the project is fundamentally trying to accomplish. Designers often confront situations where goals are ill-defined, scope shifts midstream, or success criteria remain unsettled. Rather than executing on a stable brief, they must engage in continuous reframing---interpreting shifting inputs, reconciling conflicting priorities, and redefining what the design problem even is. This form of uncertainty is epistemic, not just logistical: it reflects a lack of clarity about what should be designed, for whom, and why.

P2 describes the difficulty of building something entirely new under unclear regulatory and data conditions: \textit{``We were creating something completely new from scratch, and it was also new for the company. We didn’t know if the necessary data was available or how it would comply with European regulations, as each country has its own rules. ... It's a head scratch.''} P7 encountered similar issues due to missing information: \textit{``We were working on a project where the client provided insufficient information. We requested more information, but no answers were provided yet and the time is running out.''}

In other cases, project goals appear defined but remain conceptually unstable. P3 recalls a strategy meeting where multiple stakeholders responded vaguely to open-ended prompts: \textit{``There are many internal stakeholders like product designers, product managers, and more in that [meeting]. We explained what we were building and asked open questions. ... They did respond, but their answer was slightly weak.''} This vagueness forces designers to make their own interpretive calls about which problems are worth solving. 

Scope changes can add additional uncertainty. P6 explains how a shift in business focus introduced new technical constraints: \textit{``As we shifted the focus more to e-commerce or grocery shopping experience, uncertainty around technical limitations increased, such as how to inform the users without making them confused.''} These adjustments often require reframing earlier assumptions, deliverables, or success metrics. Finally, practitioners frequently confront tensions between business needs and user experience. P6 reflects on this conflict: \textit{``a lot of the tension comes business wise, this is a good metrics to meet''} These tensions are not incidental—they are constitutive of framing uncertainty in UX design. Designers must continually negotiate which values to prioritize, which goals to foreground, and which constraints to accept or resist. Unlike fixed problem definitions, these project framings are provisional, unstable, and subject to ongoing interpretation.

\subsubsection{Feasibility \& Constraints}
The final source of uncertainty centers on the limits of feasibility, when technical, infrastructural, or regulatory constraints come into play. At times, designs that seem complete may face unexpected blockers—from engineering hurdles to legal compliance—that force teams to revisit or even abandon their original approach. Unlike uncertainty in problem framing or planning, these issues often surface late in the process, after considerable effort has already been invested. This latency compounds their impact: designers are forced to revisit or abandon decisions they believed were resolved, undermining progress and introducing frustration, rework, and negotiation. 

Technical feasibility is a recurring concern. P2 highlights how implementation often reveals hidden labor: \textit{``After we believe a product is fully developed, developers may find that it requires an additional two months of effort, which we may not have.''} P3 shares a similar issue: \textit{''Engineers now are saying that [idea] will not work, while I'm almost on the verge of delivering the designs.''} P8 describes how sometimes entire concepts are discarded:\textit{``We developed UX wireframes and reviewed them in meetings. Then, [engineers] came back and said they couldn't implement the design.''} Post-launch or QA issues also generate new forms of uncertainty. P6 mentions how invisible system behavior led to a misaligned experience: \textit{``There were complaints where [users] searched for a certain product and it didn't exist in the search bar. But as they browsed the category page, [the searched product] was there. As this is about addressing the technical glitches for engineers, my job was more like a QC (quality control) of the overall user experience perspective.''} These episodes highlight how design quality becomes entangled with system behavior well beyond the design phase itself.

Legal and regulatory obligations also create execution challenges. P4 recalls a sudden compliance deadline: \textit{``[The company] told us that there would be a new law that came in [country]. That means that we must act on it and launch [new version] fast. If we're not compliant [to this new law], there will be issues.''} P9 discusses banking restrictions: \textit{``In [country], the concept of open banking is unfamiliar. It's hard to get the API from all the banks and sync with the bank account automatically like in other countries. So we need to find a way to work around it.''} Others described legal requirements as opaque or inaccessible. P8 explains: \textit{``[Stakeholder] was consulting with legal teams and [other decision makers], and I only got the information. When I asked what would happen to the designs, they suggested I check the contract as it included certain requirements. Turned out there were many edge cases to consider.''} In some cases, constraints were only discovered after design work had been completed. As P6 notes: \textit{``We discovered that certain aspects of the [design idea] could not be executed legally. We should have involved [the legal team] earlier, but we only realized this towards the end.''}

In these scenarios, feasibility is not just a technical property—it is a distributed, emergent condition that depends on communication, timing, and foresight. When constraints surface late or remain unstated, they render the design process retroactively uncertain, undoing assumptions that once felt secure.

\begin{table*}[ht!]
\centering
\caption{Sources of Uncertainty in UX Design Practice and Practitioner Strategies}
\label{tab:typology}
\begin{tabular}{p{3.5cm}p{6.5cm}p{6.5cm}}
\toprule
\textbf{Source of Uncertainty} & \textbf{Manifestations in Practice} & \textbf{Practitioner Strategies} \\
\midrule
\textbf{People \& Power} & 
Ambiguous or shifting project requirements; opinionated or misaligned clients; one-sided influence in decision-making. & 
Exercising contextual judgment; reframing or reinterpreting stakeholder input; negotiating boundaries and expectations. \\
\addlinespace

\textbf{Collaboration \& Coordination} & 
Misaligned team expectations; lack of clarity around deliverables and approval; ineffective communication within and across teams. & 
Building communication frameworks; clarifying roles and approval processes; establishing feedback loops and shared understanding. \\
\addlinespace

\textbf{Process \& Planning} & 
Disruption from shifting project assignments; ambiguous timelines; limited access to users or tools. & 
Setting expectations proactively; prioritizing tasks under constraint; managing stakeholder perceptions of progress. \\
\addlinespace

\textbf{Project (Re)Framing} & 
Unclear goals or problem definitions; conflicting user and business priorities; evolving project direction midstream. & 
Strategic scoping and reprioritization; stakeholder workshops to define scope; balancing user, technical, and organizational considerations. \\
\addlinespace

\textbf{Feasibility \& Constraints} & 
Late-stage technical or legal blockers; regulatory or compliance limitations; unanticipated post-launch issues. & 
Involving engineers and legal teams early; performing workaround design; conducting internal reviews and “self-testing” to anticipate problems. \\
\bottomrule
\end{tabular}
\end{table*}

\subsection{Practitioner Strategies for Navigating Uncertainty}
While the previous section identified key sources of uncertainty in UX practice, this section turns to how practitioners respond. Rather than adhering to predefined methods or linear processes, designers navigate uncertainty through framing, negotiation, experiential reasoning, and context-sensitive judgment. These responses are not merely reactive, but situated and interpretive—shaped by evolving project conditions and grounded in practical reasoning. Across the data, we observed how designers balance competing demands, reframe ambiguous problems, manage stakeholder expectations, and make trade-offs under constraint. Table~\ref{tab:judgment_strategies} summarizes these strategies, highlighting distinct forms of judgment and illustrating them with examples from our study.

\subsubsection{Framing, Judgment, and Improvisation}
In the absence of clear data or direction, UX practitioners often rely on interpretive judgment and pragmatic improvisation to move forward. These responses are not evidence of guesswork, but of situated expertise. Designers draw on experiential knowledge to assess what is `good enough,' determine when additional research is warranted, and explore alternate framings when projects stall. Participants frequently described cases where they chose not to conduct further research, trusting their internal models of the user or context. P4 explains: \textit{``At this moment, I already know the users. The flow is just a modification of an existing thing. So [user testing] was not needed since that would be a waste of resources.''} Similarly, P6 challenges the reflex to default to research: \textit{``The problem that I have a lot is you do research right away. [... for some cases,] usability testing is good if you are a customer and you should be the best judge.''} 

Practitioners also described using reframing strategies to work through conceptual blocks. P5 reflects on this practice: \textit{``I tend to dipole and go between maybe four or five different routes to solve the problems at the beginning if [the initial ones] were not working.''} Here, uncertainty is not treated as a problem to eliminate, but as a signal to explore alternate formulations of the design space. Reframing becomes a generative technique for unlocking forward momentum.

\subsubsection{Negotiating Alignment and Building Trust}
UX practitioners frequently encounter uncertainty not because a decision is technically difficult, but because the social infrastructure required to make or validate that decision is unclear. In these situations, uncertainty must be addressed through relational strategies: building alignment across teams, clarifying expectations, and establishing shared understanding about goals, roles, and process. These are not simply communication tasks—they are acts of negotiation that depend on trust, credibility, and timing.

Building communication infrastructure becomes a way to manage this uncertainty. P5 shares: \textit{``You're building a communication model with them. If you're not delivering results at the same time, you're screwing yourself because there's no trust being built up.''} Here, communication is not just about content but consistency—providing reliable feedback loops that sustain forward motion and legitimacy.


Participants also emphasized early and continuous stakeholder involvement as a strategy for reducing misalignment. P9 explains: \textit{``I will try involving the engineering team much earlier, in the phase where we have defined what we want to build. The tech team will know best whether [the idea is] feasible or not.''}. Maintaining open feedback loops is also a critical aspect of collaboration, as noted by P8 and P10, who stress the importance of continuous feedback for adapting project directions: \textit{``just lots of meetings, feedback, asking for opinions, gathering information.''}

These alignment practices are not one-time interventions. They must be sustained across the life of a project to adapt to shifts in scope, personnel, and organizational priorities. As P2 reflects: \textit{``We need a better process for approval. Because we’re stuck when we can’t proceed without a sign-off, or we can’t rework things that should have been signed off before. We had at least one workshop to re-explain what we thought was obvious... We re-explained what the process is, how it needs to be done, and reestablished roles.''} In contexts where no formal approval hierarchy exists—or where stakeholders lack UX literacy—designers must create the conditions under which their work can proceed. 


\subsubsection{Surfacing Constraints and Informing Action}
While UX practitioners often lack complete visibility into the systems and contexts they design for, they frequently use research and inquiry to surface constraints and make informed progress. These efforts are not always structured user studies; they are pragmatic moves to reduce friction, uncover blind spots, and align action with the realities of the domain. Several participants emphasized the importance of becoming domain fluent. P4 explains: \textit{``My learning is that it helps to do a lot of your homework... If you're the person that knows these [domain knowledge], you're gonna be invited to good meetings and make an impact.''} P2 echos the importance of UX designers becoming subject matter experts to influence projects beyond mere feature design, effectively reshaping business goals and structures: \textit{``We had to become experts in the same subject matter. Then we could bring expertise into how to restructure, rephrase business goals, and act more of a consultancy. So it's also about effectively designing the business structure, and not just a feature or a product.''}

Participants also used questioning strategically---to clarify the problem space, identify where friction lay, and reveal where knowledge was missing. As P2 recalls:  \textit{``I wrote down all my questions to share with the internal team to filter out the questions that might be my own ignorance. [...] Once we have common questions that we can't answer, those probably are the questions worth asking. We organized a series of meetings [with the client] to pinpoint the route and also the problem statement. Turned out, frictions were within the organization and the project specifically. Once you have the problem statement and the organized work, it becomes quite easy.''}. P3 similarly used a workshop to reveal unknowns: \textit{``I invited leads, product managers, and product designers from each team to do a workshop. Here, I asked about the requirements, processes, and more [about this project]. Then we were able to gather information [about the project] that we didn't know before.''} Further, when direct access to actual users is limited, participants report using open-source knowledge, such as internet searches or competitor analysis. 

In more constrained environments, participants employed lightweight or informal methods to simulate the effects of user testing. P7 explains: \textit{``If you don't have enough information, what are we going to do? Let's go to [online search] to take a look. So we can have an idea [about the project], about [users], about competitors.''}. In other cases, practitioners drew on their own familiarity with the product to perform self-testing, as described by P6: \textit{``We didn't conduct many tests with users because it's too complex. Some of us took on the lens of a user because we were so used to the app. We would share our observations such as our thoughts about what if this product didn't exist and so on. ... This helped us prompt idea scenarios that might or might not happen.''}

Across these cases, research is not framed as a formal phase, but as an embedded and tactical practice. Practitioners surface just enough structure to inform next steps—often under constraints of time, access, or ambiguity. These strategies reveal how uncertainty is navigated not by eliminating complexity, but by uncovering actionable contours within it.

\subsubsection{Managing Expectations and Trade-offs}
Uncertainty in UX design often stems not from ambiguity in what needs to be done, but from the realities of constrained time, attention, and resources. Practitioners regularly face situations where multiple stakeholders demand parallel progress, timelines are compressed, or expectations exceed what is feasible within the scope of a project. In these moments, designers rely on transparent expectation-setting and trade-off negotiation to maintain alignment and credibility while keeping the project on track.

P4 describes one such moment of proactive boundary-setting: \textit{``Communication is the key here to set expectations with both to say, ‘Hey, there’s something that came up on this feature. I cannot dedicate that much time to you. Can I give you 20\% of my week? And then next week I can do more.’ It’s also managing deliverables.''} This quote exemplifies a strategy of openly communicating limitations—not to withdraw from responsibility, but to sustain trust by providing a rationale and a pathway forward. P5 describes how prioritization becomes a negotiation that blends design goals with business imperatives: \textit{``What are the top problems, and how do we rank them? ... If you don’t consider the financial impact, that’s also a problem.''} Here, managing uncertainty requires not just technical judgment, but interpretive and value-laden decision-making across multiple competing metrics.

\begin{table*}[ht]
\centering
\caption{Strategies for Navigating Uncertainty in UX Practice}
\label{tab:judgment_strategies}
\begin{tabular}{p{3.5cm}p{6.7cm}p{6.6cm}}
\toprule
\textbf{Strategy Type} & \textbf{Description} & \textbf{Illustrative Practices} \\
\midrule

\textbf{Framing, Sacrifice, and Improvisation} & 
Reframing problems, improvising when data is lacking, and sacrificing ideal solutions to move forward. & 
- P5 explores alternative framings when stuck. \newline
- P4 chooses not to run usability testing due to scope and effort constraints. \\

\addlinespace

\textbf{Negotiating Alignment and Building Trust} & 
Clarifying expectations and roles, and cultivating stakeholder confidence over time. & 
- P2 re-explains sign-off processes to align stakeholders. \newline
- P5 builds trust through consistent communication. \\

\addlinespace

\textbf{Surfacing Constraints and Informing Action} & 
Identifying missing knowledge or barriers to guide feasible decisions. & 
- P2 compiles and filters key questions to prioritize. \newline
- P7 conducts open-source research to simulate user feedback. \\

\addlinespace

\textbf{Managing Expectations and Trade-offs} & 
Setting boundaries and negotiating deliverables under time and resource constraints. & 
- P4 allocates only 20\% of time to a task and renegotiates future involvement. \newline
- P5 prioritizes design work based on business impact. \\
\bottomrule
\end{tabular}
\end{table*}

\section{Discussion}
Our findings show that uncertainty is not merely a background condition of UX work---it is a pervasive and structuring feature of practice. While prior literature has emphasized the ill-structured nature of design problems \cite{simon_structure_1973, rittel_dilemmas_1973}, and the reflective framing required to navigate them \cite{schon_reflective_1983}, our study extends these insights by examining how UX practitioners experience and manage uncertainty in daily practice. Practitioners do not eliminate uncertainty; they work through it—often improvisationally—using strategies grounded in situated judgment, communication, and experiential reasoning.

\subsection{Uncertainty as a Structuring Condition of UX Design}
Uncertainty in UX practice is not simply a knowledge gap to be resolved through better research or process. It arises from layered contexts in which UX work unfolds: stakeholder misalignment, ambiguous goals, fragmented timelines, and shifting constraints. The five sources we surfaced---spanning people and power, collaboration and coordination, process and planning, project (re)framing, and feasibility---highlight how uncertainty is embedded in the institutional and temporal fabric of UX practice. While we organize these findings using five types, this structure is intended as a pragmatic lens rather than a fixed typology. In practice, these sources of uncertainty often intersect and entangle. Our categories aim to foreground recurring tensions while acknowledging the overlapping, situated nature of UX work.


These challenges are not unique to UX design, but the user-centered, cross-functional nature of UX roles introduces particular complications. Designers must often manage both external stakeholders (e.g., clients, users) and internal collaborators (e.g., engineers, product managers), all while maintaining coherence across tools, timelines, and value systems. As prior work has noted, a degree of uncertainty can even be beneficial. Daalhuizen et al. \cite{daalhuizen_dealing_2009} and Garvey and Childs \cite{garvey_design_2016} argue that moderate levels of uncertainty can act as a catalyst for creativity by allowing room for exploration and reinterpretation. Our study supports this view: practitioners do not always seek to eliminate uncertainty, but rather to work productively within it.


In some cases, as P4 illustrates, strategic domain knowledge becomes a way to reduce the impact of uncertainty and gain credibility in ambiguous environments. \textit{``If you're the person that knows these things [domain knowledge], you're gonna be invited to good meetings and make an impact.''} This finding contrasts with prior work suggesting that UX designers do not necessarily see technical or cross-domain knowledge (e.g., machine learning) as central to their craft \cite{yang_investigating_2018}. In our data, the acquisition of contextual knowledge is not just instrumental, it is a condition for participating in key conversations and influencing problem framing.

Uncertainty in UX design is not an episodic disruption; rather, it is a structuring condition. It arises not from a lack of method, but from the nature of the environments in which UX design is embedded: complex, evolving, politically charged, and resource-constrained. Addressing these uncertainties one at a time is rarely sufficient; instead, designers must develop strategies for navigating uncertainty as a condition of practice itself.


\subsection{Situated Judgment through Framing, Sacrifice, and Improvisation}
UX practitioners do not respond to uncertainty by following prescriptive methods. Instead, they rely on situated judgment---contextual, embodied, and evolving knowledge drawn from practice. While popular design models like the Double Diamond \cite{dam_what_2021} may provide scaffolding for novices, they do not capture the practices of experienced designers, who navigating uncertainty through iterative reasoning, strategic reframing, and adaptive compromise. As Cross \cite{cross_design_2023} notes, design thinking in practice increasingly functions as strategic, adaptive intelligence for addressing complex, wicked problems. 

This adaptive intelligence involves multiple forms of situated judgment. Practitioners make real-time decisions with incomplete knowledge, ambiguous constraints, and multiple stakeholders, often without the opportunity to pause for formal research or validation. These moments reflect Schön's ``reflection-in-action'' \cite{schon_reflective_1983}, and what Nelson and Stolterman theorized as a rich set of judgment types in design practice \cite{nelson_design_2002}, including appreciative (deciding what to prioritize), navigational (choosing direction amid uncertainty), and deliberated off-hand (actions shaped by experience and intuition) judgments. All three were visible in our findings. For example, P6 made a navigational judgment in choosing usability testing over new research, while P3 demonstrated appreciative judgment by prioritizing low-fidelity over high-fidelity wireframes. These moments exemplify how design decisions emerge not from fixed rules, but from adaptive negotiation of constraints.

We also observed two additional forms of judgment that extend existing theory: \textit{temporal} and \textit{sacrificial} judgments. Temporal judgment involves anticipating how timing affects decisions and acting accordingly. P1 shows this through awareness of pacing: \textit{``We only have five weeks left... we won’t be able to schedule or synthesize the interviews.''} P2 adds a strategic layer by weighing time against value: \textit{``How long does it take... versus how much money can we anticipate?''} P4 reflects on poor temporal judgment: \textit{``I didn’t do my stuff on time... delaying the project.''} Together, these cases underscore the importance of temporal judgment. Sacrificial judgment, by contrast, entails the deliberate abandonment of one goal in service of another. While this concept has appeared in resilience engineering literature as a way to describe goal trade-offs under pressure \cite{woods_essential_2006}, we observed it clearly in design settings. P3 decided to: \textit{``let go of proper user research and go ahead with a few customer interactions.''} P4 similarly decided to skip usability testing due to effort constraints and limited expected impact. These judgments are not failures of method; they are evidence of experienced designers adapting to constraint---evidence that judgment, not process adherence, is often what carries design work forward.

\subsection{Cultivating Judgment and Adaptability in UX Education} 

Our findings point to judgment and adaptability, not just process literacy or tool proficiency, as central capacities in UX practice. Yet current UX educational models often fall short in preparing students for the interpretive, relational, and context-sensitive nature of real-world design work. As Branch et al. \cite{branch_user_2021} and Kou and Gray \cite{kou_practice-led_2019} observe, many academic UX programs lag behind the evolving needs of the profession. The result is a persistent disconnect between the structured scenarios emphasized in coursework and the complex, uncertain realities of practice.

Educators have called for more immersive, practice-linked learning environments---e.g., exposing students to framing problems across diverse contexts \cite{parsons_developing_2023}, and integrating client-facing, collaborative projects to simulate real-world ambiguity and complexity \cite{macdonald_assessing_2017, vorvoreanu_advancing_2017}. Studio-based models, when designed to foreground reflection and responsiveness, align closely with Schön’s concept of the `reflective practicum' \cite{schon_educating_1987}, which emphasize learning through the complex doing of professional judgment. Gray and Parsons \cite{gray_building_2023} extend this view by arguing that teaching design methods must go beyond procedural fluency to emphasize adaptive mastery---the ability to modify, subvert, or resist method prescriptions based on local project conditions.

This emphasis on cultivating judgment is supported by recent work in instructional design. Demiral-Uzan and Boling \cite{demiral-uzan_instructional_2024} show that judgment develops through real-world constraints, critique, and iterative decision-making. This aligns with our finding that UX practitioners interpret and improvise under pressure. Teaching such competencies requires moving beyond standard methods to help students practice reframing, managing expectations, and negotiating scope. Communication, alignment, and negotiation must also be taught as core design skills \cite{macdonald_how_2022, shukla_communication_2024}.

This imperative also connects to emerging concerns around leadership in UX. Meyer and Norman \cite{meyer_changing_2020} argue that the absence of UX professionals in leadership roles is tied not to a lack of tactical skill, but to gaps in broader design judgment, critical communication, and organizational awareness. If UX education aims to prepare designers for influence—not just implementation—it must cultivate the ability to frame problems, adapt under pressure, and reason through uncertainty. This may require new curricular structures, but also new priorities: emphasizing strategic judgment over procedural fidelity, and adaptability over methodological purity. Preparing UX students to navigate uncertainty means preparing them to judge, not just to execute. 

\section{Limitations}

The study included a relatively small sample of 10 UX practitioners from different regions, which may limit its ability to capture the full spectrum of uncertainties across diverse domains and contexts. To report a broad range of uncertainties while acknowledging variability by experience level, we included participants with 2 to 11+ years of experience. However, we did not examine how expertise shapes perceptions of uncertainty---an opportunity for future research. Nor did we explore differences across sectors, such as healthcare, finance, or e-commerce. Several participants from Europe and Asia-Pacific were non-native English speakers, which may have limited their ability to fully articulate experiences. We also did not consider uncertainties introduced by emerging technologies like generative AI, which can subtly reshape design roles and responsibilities \cite{shukla_-skilling_2025}. Our findings may not be transferable to UX work in domains with different regulatory, organizational, or user dynamics from those of our participants. We see our findings as likely transferable, but caution is always warranted.

All data were self-reported through diary entries and interviews, reflecting individual perspective on uncertainty. Although diaries were completed weekly, recall bias remains a concern due to delays between events and documentation. We also did not collect any artifacts---such as briefs, user stories, wireframes, or final deliverables---so our analysis is based entirely on participants' recollections. Additionally, the four-week period may not have captured uncertainty across all project phases. While interviews explored diary entries in greater depth, participants' reflections may not fully reflect how they experienced uncertainty in real-time.

\section{Conclusion}
This study contributes a grounded account of how UX designers experience and navigate uncertainty in their everyday practice. Through a combination of diary studies and interviews, we identified five recurring domains of uncertainty and four strategic categories of situated judgment that practitioners use to respond. Our findings show that uncertainty is deeply intertwined with the dynamic, real-world contexts of design, where evolving project goals, stakeholder demands, and technical constraints add complexity to the practice. While this study lays a foundation for understanding uncertainty in UX design practice, further research is needed to refine these categories, especially given the rapid evolution of technologies. Additionally, studies on the role of judgment-making in UX design could enhance our understanding of how designers manage uncertainty.

\begin{acks}
This work was supported by NSF award \#2146228.
\end{acks}

\bibliographystyle{ACM-Reference-Format}
\bibliography{references}


\begin{thebibliography}{68}


\ifx \showCODEN    \undefined \def \showCODEN     #1{\unskip}     \fi
\ifx \showISBNx    \undefined \def \showISBNx     #1{\unskip}     \fi
\ifx \showISBNxiii \undefined \def \showISBNxiii  #1{\unskip}     \fi
\ifx \showISSN     \undefined \def \showISSN      #1{\unskip}     \fi
\ifx \showLCCN     \undefined \def \showLCCN      #1{\unskip}     \fi
\ifx \shownote     \undefined \def \shownote      #1{#1}          \fi
\ifx \showarticletitle \undefined \def \showarticletitle #1{#1}   \fi
\ifx \showURL      \undefined \def \showURL       {\relax}        \fi
\providecommand\bibfield[2]{#2}
\providecommand\bibinfo[2]{#2}
\providecommand\natexlab[1]{#1}
\providecommand\showeprint[2][]{arXiv:#2}

\bibitem[Bar-Anan et~al\mbox{.}(2009)]%
        {bar-anan_feeling_2009}
\bibfield{author}{\bibinfo{person}{Yoav Bar-Anan}, \bibinfo{person}{Timothy~D. Wilson}, {and} \bibinfo{person}{Daniel~T. Gilbert}.} \bibinfo{year}{2009}\natexlab{}.
\newblock \showarticletitle{The feeling of uncertainty intensifies affective reactions}.
\newblock \bibinfo{journal}{\emph{Emotion}} \bibinfo{volume}{9}, \bibinfo{number}{1} (\bibinfo{year}{2009}), \bibinfo{pages}{123--127}.
\newblock
\showISSN{1931-1516}
\href{https://doi.org/10.1037/a0014607}{doi:\nolinkurl{10.1037/a0014607}}
\newblock
\shownote{Place: US Publisher: American Psychological Association}.


\bibitem[Branch et~al\mbox{.}(2021)]%
        {branch_user_2021}
\bibfield{author}{\bibinfo{person}{James Branch}, \bibinfo{person}{Christopher~J. Parker}, {and} \bibinfo{person}{Mark Evans}.} \bibinfo{year}{2021}\natexlab{}.
\newblock \showarticletitle{Do {User} {Experience} ({UX}) {Design} {Courses} {Meet} {Industry}’s {Needs}? {Analysing} {UX} {Degrees} and {Job} {Adverts}}.
\newblock \bibinfo{journal}{\emph{The Design Journal}} \bibinfo{volume}{24}, \bibinfo{number}{4} (\bibinfo{date}{July} \bibinfo{year}{2021}), \bibinfo{pages}{631--652}.
\newblock
\showISSN{1460-6925}
\href{https://doi.org/10.1080/14606925.2021.1930935}{doi:\nolinkurl{10.1080/14606925.2021.1930935}}
\newblock
\shownote{Publisher: Routledge \_eprint: https://doi.org/10.1080/14606925.2021.1930935}.


\bibitem[Buchanan(1992)]%
        {buchanan_wicked_1992}
\bibfield{author}{\bibinfo{person}{Richard Buchanan}.} \bibinfo{year}{1992}\natexlab{}.
\newblock \showarticletitle{Wicked {Problems} in {Design} {Thinking}}.
\newblock \bibinfo{journal}{\emph{Design Issues}} (\bibinfo{year}{1992}).
\newblock


\bibitem[Cajander et~al\mbox{.}(2022)]%
        {cajander_ux_2022}
\bibfield{author}{\bibinfo{person}{Åsa Cajander}, \bibinfo{person}{Marta Larusdottir}, {and} \bibinfo{person}{Johannes~L. Geiser}.} \bibinfo{year}{2022}\natexlab{}.
\newblock \showarticletitle{{UX} professionals’ learning and usage of {UX} methods in agile}.
\newblock \bibinfo{journal}{\emph{Information and Software Technology}}  \bibinfo{volume}{151} (\bibinfo{date}{Nov.} \bibinfo{year}{2022}), \bibinfo{pages}{107005}.
\newblock
\showISSN{0950-5849}
\href{https://doi.org/10.1016/j.infsof.2022.107005}{doi:\nolinkurl{10.1016/j.infsof.2022.107005}}


\bibitem[Crescenzi et~al\mbox{.}(2016)]%
        {crescenzi_impacts_2016}
\bibfield{author}{\bibinfo{person}{Anita Crescenzi}, \bibinfo{person}{Diane Kelly}, {and} \bibinfo{person}{Leif Azzopardi}.} \bibinfo{year}{2016}\natexlab{}.
\newblock \showarticletitle{Impacts of {Time} {Constraints} and {System} {Delays} on {User} {Experience}}. In \bibinfo{booktitle}{\emph{Proceedings of the 2016 {ACM} on {Conference} on {Human} {Information} {Interaction} and {Retrieval}}}. \bibinfo{publisher}{ACM}, \bibinfo{address}{Carrboro North Carolina USA}, \bibinfo{pages}{141--150}.
\newblock
\showISBNx{978-1-4503-3751-9}
\href{https://doi.org/10.1145/2854946.2854976}{doi:\nolinkurl{10.1145/2854946.2854976}}


\bibitem[Cross(2023)]%
        {cross_design_2023}
\bibfield{author}{\bibinfo{person}{Nigel Cross}.} \bibinfo{year}{2023}\natexlab{}.
\newblock \bibinfo{booktitle}{\emph{Design {Thinking}: {Understanding} {How} {Designers} {Think} and {Work}}}.
\newblock \bibinfo{publisher}{Bloomsbury Publishing}.
\newblock
\showISBNx{978-1-350-30507-6}


\bibitem[Cross et~al\mbox{.}(1981)]%
        {cross_design_1981}
\bibfield{author}{\bibinfo{person}{Nigel Cross}, \bibinfo{person}{John Naughton}, {and} \bibinfo{person}{David Walker}.} \bibinfo{year}{1981}\natexlab{}.
\newblock \showarticletitle{Design method and scientific method}.
\newblock \bibinfo{journal}{\emph{Design Studies}} \bibinfo{volume}{2}, \bibinfo{number}{4} (\bibinfo{date}{Oct.} \bibinfo{year}{1981}), \bibinfo{pages}{195--201}.
\newblock
\showISSN{0142694X}
\href{https://doi.org/10.1016/0142-694X(81)90050-8}{doi:\nolinkurl{10.1016/0142-694X(81)90050-8}}


\bibitem[Daalhuizen et~al\mbox{.}(2009)]%
        {daalhuizen_dealing_2009}
\bibfield{author}{\bibinfo{person}{Jaap Daalhuizen}, \bibinfo{person}{Petra Badke-Schaub}, {and} \bibinfo{person}{Batill Stephen}.} \bibinfo{year}{2009}\natexlab{}.
\newblock \showarticletitle{Dealing with uncertainty in design practice: issues for designer-centered methodology}.
\newblock In \bibinfo{booktitle}{\emph{{ICED} 09 - {The} 17th {International} {Conference} on {Engineering} {Design}, {Vol} 9: {Human} {Behavior} in {Design}}}. Number Vol. 9. \bibinfo{publisher}{Design Society}, \bibinfo{address}{Glasgow}.
\newblock
\showISBNx{978-1-904670-13-1}


\bibitem[Dam and Siang(2021)]%
        {dam_what_2021}
\bibfield{author}{\bibinfo{person}{Rikke~Friis Dam} {and} \bibinfo{person}{Teo~Yu Siang}.} \bibinfo{year}{2021}\natexlab{}.
\newblock \showarticletitle{What is {Design} {Thinking} and {Why} {Is} {It} {So} {Popular}?}
\newblock \bibinfo{journal}{\emph{Interaction Design Foundation}} (\bibinfo{year}{2021}).
\newblock


\bibitem[de~Weck et~al\mbox{.}(2007)]%
        {de_weck_classification_2007}
\bibfield{author}{\bibinfo{person}{Olivier de Weck}, \bibinfo{person}{Claudia Eckert}, {and} \bibinfo{person}{John Clarkson}.} \bibinfo{year}{2007}\natexlab{}.
\newblock \showarticletitle{A classification of uncertainty for early product and system design}.
\newblock \bibinfo{journal}{\emph{DS 42: Proceedings of ICED 2007, the 16th International Conference on Engineering Design, Paris, France}} (\bibinfo{year}{2007}), \bibinfo{pages}{159--160}.
\newblock


\bibitem[Demiral-Uzan and Boling(2024)]%
        {demiral-uzan_instructional_2024}
\bibfield{author}{\bibinfo{person}{Muruvvet Demiral-Uzan} {and} \bibinfo{person}{Elizabeth Boling}.} \bibinfo{year}{2024}\natexlab{}.
\newblock \showarticletitle{Instructional design students’ design judgment development}.
\newblock \bibinfo{journal}{\emph{Educational technology research and development}} \bibinfo{volume}{72}, \bibinfo{number}{3} (\bibinfo{date}{June} \bibinfo{year}{2024}), \bibinfo{pages}{1813--1849}.
\newblock
\showISSN{1042-1629, 1556-6501}
\href{https://doi.org/10.1007/s11423-024-10361-1}{doi:\nolinkurl{10.1007/s11423-024-10361-1}}


\bibitem[Dorst(2003)]%
        {dorst_exploring_2003}
\bibfield{author}{\bibinfo{person}{Kees Dorst}.} \bibinfo{year}{2003}\natexlab{}.
\newblock \showarticletitle{Exploring the structure of design problems}.
\newblock \bibinfo{journal}{\emph{Internation conference on Engineering Design}} (\bibinfo{year}{2003}).
\newblock


\bibitem[Dorst(2004)]%
        {dorst_problem_2004}
\bibfield{author}{\bibinfo{person}{Kees Dorst}.} \bibinfo{year}{2004}\natexlab{}.
\newblock \showarticletitle{The {Problem} of {Design} {Problems}}.
\newblock \bibinfo{journal}{\emph{Journal of design research}} (\bibinfo{year}{2004}), \bibinfo{pages}{4(2), 185--196}.
\newblock


\bibitem[Dorst(2011)]%
        {dorst_core_2011}
\bibfield{author}{\bibinfo{person}{Kees Dorst}.} \bibinfo{year}{2011}\natexlab{}.
\newblock \showarticletitle{The core of ‘design thinking’ and its application}.
\newblock \bibinfo{journal}{\emph{Design Studies}} \bibinfo{volume}{32}, \bibinfo{number}{6} (\bibinfo{date}{Nov.} \bibinfo{year}{2011}), \bibinfo{pages}{521--532}.
\newblock
\showISSN{0142694X}
\href{https://doi.org/10.1016/j.destud.2011.07.006}{doi:\nolinkurl{10.1016/j.destud.2011.07.006}}


\bibitem[Dorst(2015)]%
        {dorst_frame_2015}
\bibfield{author}{\bibinfo{person}{Kees Dorst}.} \bibinfo{year}{2015}\natexlab{}.
\newblock \showarticletitle{Frame {Creation} and {Design} in the {Expanded} {Field}}.
\newblock \bibinfo{journal}{\emph{She Ji: The Journal of Design, Economics, and Innovation}} \bibinfo{volume}{1}, \bibinfo{number}{1} (\bibinfo{year}{2015}), \bibinfo{pages}{22--33}.
\newblock
\showISSN{24058726}
\href{https://doi.org/10.1016/j.sheji.2015.07.003}{doi:\nolinkurl{10.1016/j.sheji.2015.07.003}}


\bibitem[Dorst and Cross(2001)]%
        {dorst_creativity_2001}
\bibfield{author}{\bibinfo{person}{Kees Dorst} {and} \bibinfo{person}{Nigel Cross}.} \bibinfo{year}{2001}\natexlab{}.
\newblock \showarticletitle{Creativity in the design process: co-evolution of problem–solution}.
\newblock \bibinfo{journal}{\emph{Design Studies}} \bibinfo{volume}{22}, \bibinfo{number}{5} (\bibinfo{date}{Sept.} \bibinfo{year}{2001}), \bibinfo{pages}{425--437}.
\newblock
\showISSN{0142694X}
\href{https://doi.org/10.1016/S0142-694X(01)00009-6}{doi:\nolinkurl{10.1016/S0142-694X(01)00009-6}}


\bibitem[Fedosov et~al\mbox{.}(2021)]%
        {fedosov_challenges_2021}
\bibfield{author}{\bibinfo{person}{Anton Fedosov}, \bibinfo{person}{Daniel Boos}, \bibinfo{person}{Susanne Schmidt-Rauch}, \bibinfo{person}{Jarno Ojala}, {and} \bibinfo{person}{Myriam Lewkowicz}.} \bibinfo{year}{2021}\natexlab{}.
\newblock \showarticletitle{Challenges of transferring {UX} designs and insights across products and services}.
\newblock \bibinfo{journal}{\emph{ECSCW}} (\bibinfo{year}{2021}).
\newblock


\bibitem[Fereday and Muir-Cochrane(2006)]%
        {fereday_demonstrating_2006}
\bibfield{author}{\bibinfo{person}{Jennifer Fereday} {and} \bibinfo{person}{Eimear Muir-Cochrane}.} \bibinfo{year}{2006}\natexlab{}.
\newblock \showarticletitle{Demonstrating {Rigor} {Using} {Thematic} {Analysis}: {A} {Hybrid} {Approach} of {Inductive} and {Deductive} {Coding} and {Theme} {Development}}.
\newblock \bibinfo{journal}{\emph{International Journal of Qualitative Methods}} \bibinfo{volume}{5}, \bibinfo{number}{1} (\bibinfo{year}{2006}), \bibinfo{pages}{80--92}.
\newblock
\showISSN{1609-4069}
\href{https://doi.org/10.1177/160940690600500107}{doi:\nolinkurl{10.1177/160940690600500107}}


\bibitem[Frishammar et~al\mbox{.}(2011)]%
        {frishammar_beyond_2011}
\bibfield{author}{\bibinfo{person}{J. Frishammar}, \bibinfo{person}{H. Floren}, {and} \bibinfo{person}{J. Wincent}.} \bibinfo{year}{2011}\natexlab{}.
\newblock \showarticletitle{Beyond {Managing} {Uncertainty}: {Insights} {From} {Studying} {Equivocality} in the {Fuzzy} {Front} {End} of {Product} and {Process} {Innovation} {Projects}}.
\newblock \bibinfo{journal}{\emph{IEEE Transactions on Engineering Management}} \bibinfo{volume}{58}, \bibinfo{number}{3} (\bibinfo{date}{Aug.} \bibinfo{year}{2011}), \bibinfo{pages}{551--563}.
\newblock
\showISSN{0018-9391, 1558-0040}
\href{https://doi.org/10.1109/TEM.2010.2095017}{doi:\nolinkurl{10.1109/TEM.2010.2095017}}


\bibitem[Garvey and Childs(2016)]%
        {garvey_design_2016}
\bibfield{author}{\bibinfo{person}{Bruce Garvey} {and} \bibinfo{person}{Peter Childs}.} \bibinfo{year}{2016}\natexlab{}.
\newblock \showarticletitle{Design as an {Unstructured} {Problem}: {New} {Methods} to {Help} {Reduce} {Uncertainty}—{A} {Practitioner} {Perspective}}.
\newblock In \bibinfo{booktitle}{\emph{Impact of {Design} {Research} on {Industrial} {Practice}: {Tools}, {Technology}, and {Training}}}, \bibfield{editor}{\bibinfo{person}{Amaresh Chakrabarti} {and} \bibinfo{person}{Udo Lindemann}} (Eds.). \bibinfo{publisher}{Springer International Publishing}, \bibinfo{address}{Cham}, \bibinfo{pages}{333--352}.
\newblock
\showISBNx{978-3-319-19449-3}
\href{https://doi.org/10.1007/978-3-319-19449-3_22}{doi:\nolinkurl{10.1007/978-3-319-19449-3_22}}


\bibitem[Goodman et~al\mbox{.}(2011)]%
        {goodman_understanding_2011}
\bibfield{author}{\bibinfo{person}{Elizabeth Goodman}, \bibinfo{person}{Erik Stolterman}, {and} \bibinfo{person}{Ron Wakkary}.} \bibinfo{year}{2011}\natexlab{}.
\newblock \showarticletitle{Understanding interaction design practices}. In \bibinfo{booktitle}{\emph{Proceedings of the 2011 annual conference on {Human} factors in computing systems - {CHI} '11}}. \bibinfo{publisher}{ACM Press}, \bibinfo{address}{Vancouver, BC, Canada}, \bibinfo{pages}{1061}.
\newblock
\showISBNx{978-1-4503-0228-9}
\href{https://doi.org/10.1145/1978942.1979100}{doi:\nolinkurl{10.1145/1978942.1979100}}


\bibitem[Gray(2016)]%
        {gray_its_2016}
\bibfield{author}{\bibinfo{person}{Colin~M. Gray}.} \bibinfo{year}{2016}\natexlab{}.
\newblock \showarticletitle{"{It}'s {More} of a {Mindset} {Than} a {Method}": {UX} {Practitioners}' {Conception} of {Design} {Methods}}. In \bibinfo{booktitle}{\emph{Proceedings of the 2016 {CHI} {Conference} on {Human} {Factors} in {Computing} {Systems}}}. \bibinfo{publisher}{ACM}, \bibinfo{address}{San Jose California USA}, \bibinfo{pages}{4044--4055}.
\newblock
\showISBNx{978-1-4503-3362-7}
\href{https://doi.org/10.1145/2858036.2858410}{doi:\nolinkurl{10.1145/2858036.2858410}}


\bibitem[Gray et~al\mbox{.}(2024)]%
        {gray_legal_2024}
\bibfield{author}{\bibinfo{person}{Colin~M. Gray}, \bibinfo{person}{Ritika Gairola}, \bibinfo{person}{Nayah Boucaud}, \bibinfo{person}{Maliha Hashmi}, \bibinfo{person}{Shruthi~Sai Chivukula}, \bibinfo{person}{Ambika~R Menon}, {and} \bibinfo{person}{Ja-Nae Duane}.} \bibinfo{year}{2024}\natexlab{}.
\newblock \showarticletitle{Legal {Trouble}?: {UX} {Practitioners}' {Engagement} with {Law} and {Regulation}}. In \bibinfo{booktitle}{\emph{Designing {Interactive} {Systems} {Conference}}}. \bibinfo{publisher}{ACM}, \bibinfo{address}{IT University of Copenhagen Denmark}, \bibinfo{pages}{106--110}.
\newblock
\showISBNx{9798400706325}
\href{https://doi.org/10.1145/3656156.3663698}{doi:\nolinkurl{10.1145/3656156.3663698}}


\bibitem[Gray and Parsons(2023)]%
        {gray_building_2023}
\bibfield{author}{\bibinfo{person}{Colin~M. Gray} {and} \bibinfo{person}{Paul~C. Parsons}.} \bibinfo{year}{2023}\natexlab{}.
\newblock \showarticletitle{Building {Student} {Capacity} to {Engage} with {Design} {Methods}}. In \bibinfo{booktitle}{\emph{Proceedings of the 5th {Annual} {Symposium} on {HCI} {Education}}}. \bibinfo{publisher}{ACM}, \bibinfo{address}{Hamburg Germany}, \bibinfo{pages}{19--22}.
\newblock
\showISBNx{9798400707377}
\href{https://doi.org/10.1145/3587399.3587415}{doi:\nolinkurl{10.1145/3587399.3587415}}


\bibitem[Gray et~al\mbox{.}(2023)]%
        {gray_dark_2023}
\bibfield{author}{\bibinfo{person}{Colin~M. Gray}, \bibinfo{person}{Cristiana~Teixeira Santos}, \bibinfo{person}{Nicole Tong}, \bibinfo{person}{Thomas Mildner}, \bibinfo{person}{Arianna Rossi}, \bibinfo{person}{Johanna~T. Gunawan}, {and} \bibinfo{person}{Caroline Sinders}.} \bibinfo{year}{2023}\natexlab{}.
\newblock \showarticletitle{Dark {Patterns} and the {Emerging} {Threats} of {Deceptive} {Design} {Practices}}. In \bibinfo{booktitle}{\emph{Extended {Abstracts} of the 2023 {CHI} {Conference} on {Human} {Factors} in {Computing} {Systems}}}. \bibinfo{publisher}{ACM}, \bibinfo{address}{Hamburg Germany}, \bibinfo{pages}{1--4}.
\newblock
\showISBNx{978-1-4503-9422-2}
\href{https://doi.org/10.1145/3544549.3583173}{doi:\nolinkurl{10.1145/3544549.3583173}}


\bibitem[Griffin and Grote(2020)]%
        {griffin_when_2020}
\bibfield{author}{\bibinfo{person}{Mark~A. Griffin} {and} \bibinfo{person}{Gudela Grote}.} \bibinfo{year}{2020}\natexlab{}.
\newblock \showarticletitle{When {Is} {More} {Uncertainty} {Better}? {A} {Model} of {Uncertainty} {Regulation} and {Effectiveness}}.
\newblock \bibinfo{journal}{\emph{Academy of Management Review}} \bibinfo{volume}{45}, \bibinfo{number}{4} (\bibinfo{date}{Oct.} \bibinfo{year}{2020}), \bibinfo{pages}{745--765}.
\newblock
\showISSN{0363-7425, 1930-3807}
\href{https://doi.org/10.5465/amr.2018.0271}{doi:\nolinkurl{10.5465/amr.2018.0271}}


\bibitem[Jones and Thoma(2019)]%
        {jones_determinants_2019}
\bibfield{author}{\bibinfo{person}{Alexander Jones} {and} \bibinfo{person}{Volker Thoma}.} \bibinfo{year}{2019}\natexlab{}.
\newblock \showarticletitle{Determinants for {Successful} {Agile} {Collaboration} between {UX} {Designers} and {Software} {Developers} in a {Complex} {Organisation}}.
\newblock \bibinfo{journal}{\emph{International Journal of Human–Computer Interaction}} (\bibinfo{date}{Dec.} \bibinfo{year}{2019}).
\newblock
\showISSN{1044-7318}
\urldef\tempurl%
\url{https://www.tandfonline.com/doi/full/10.1080/10447318.2019.1587856}
\showURL{%
\tempurl}
\newblock
\shownote{Publisher: Taylor \& Francis}.


\bibitem[Klein et~al\mbox{.}(1989)]%
        {klein_critical_1989}
\bibfield{author}{\bibinfo{person}{G.A. Klein}, \bibinfo{person}{R. Calderwood}, {and} \bibinfo{person}{D. MacGregor}.} \bibinfo{year}{1989}\natexlab{}.
\newblock \showarticletitle{Critical decision method for eliciting knowledge}.
\newblock \bibinfo{journal}{\emph{IEEE Transactions on Systems, Man, and Cybernetics}} \bibinfo{volume}{19}, \bibinfo{number}{3} (\bibinfo{date}{June} \bibinfo{year}{1989}), \bibinfo{pages}{462--472}.
\newblock
\showISSN{00189472}
\href{https://doi.org/10.1109/21.31053}{doi:\nolinkurl{10.1109/21.31053}}


\bibitem[Kou and Gray(2019)]%
        {kou_practice-led_2019}
\bibfield{author}{\bibinfo{person}{Yubo Kou} {and} \bibinfo{person}{Colin~M. Gray}.} \bibinfo{year}{2019}\natexlab{}.
\newblock \showarticletitle{A {Practice}-{Led} {Account} of the {Conceptual} {Evolution} of {UX} {Knowledge}}. In \bibinfo{booktitle}{\emph{Proceedings of the 2019 {CHI} {Conference} on {Human} {Factors} in {Computing} {Systems}}}. \bibinfo{publisher}{ACM}, \bibinfo{address}{Glasgow Scotland Uk}, \bibinfo{pages}{1--13}.
\newblock
\showISBNx{978-1-4503-5970-2}
\href{https://doi.org/10.1145/3290605.3300279}{doi:\nolinkurl{10.1145/3290605.3300279}}


\bibitem[Lipshitz and Strauss(1997)]%
        {lipshitz_coping_1997}
\bibfield{author}{\bibinfo{person}{Raanan Lipshitz} {and} \bibinfo{person}{Orna Strauss}.} \bibinfo{year}{1997}\natexlab{}.
\newblock \showarticletitle{Coping with {Uncertainty}: {A} {Naturalistic} {Decision}-{Making} {Analysis}}.
\newblock \bibinfo{journal}{\emph{Organizational Behavior and Human Decision Processes}} \bibinfo{volume}{69}, \bibinfo{number}{2} (\bibinfo{date}{Feb.} \bibinfo{year}{1997}), \bibinfo{pages}{149--163}.
\newblock
\showISSN{0749-5978}
\href{https://doi.org/10.1006/obhd.1997.2679}{doi:\nolinkurl{10.1006/obhd.1997.2679}}


\bibitem[MacDonald et~al\mbox{.}(2022)]%
        {macdonald_how_2022}
\bibfield{author}{\bibinfo{person}{Craig~M. MacDonald}, \bibinfo{person}{Emma~J. Rose}, {and} \bibinfo{person}{Cynthia Putnam}.} \bibinfo{year}{2022}\natexlab{}.
\newblock \showarticletitle{How, {Why}, and with {Whom} {Do} {User} {Experience} ({UX}) {Practitioners} {Communicate}? {Implications} for {HCI} {Education}}.
\newblock \bibinfo{journal}{\emph{International Journal of Human–Computer Interaction}} \bibinfo{volume}{38}, \bibinfo{number}{15} (\bibinfo{date}{Sept.} \bibinfo{year}{2022}), \bibinfo{pages}{1422--1439}.
\newblock
\showISSN{1044-7318, 1532-7590}
\href{https://doi.org/10.1080/10447318.2021.2002050}{doi:\nolinkurl{10.1080/10447318.2021.2002050}}


\bibitem[MacDonald and Rozaklis(2017)]%
        {macdonald_assessing_2017}
\bibfield{author}{\bibinfo{person}{Craig~M. MacDonald} {and} \bibinfo{person}{Lillian Rozaklis}.} \bibinfo{year}{2017}\natexlab{}.
\newblock \showarticletitle{Assessing the implementation of authentic, client-facing student projects in user experience ({UX}) education: {Insights} from multiple stakeholders}.
\newblock \bibinfo{journal}{\emph{Proceedings of the Association for Information Science and Technology}} \bibinfo{volume}{54}, \bibinfo{number}{1} (\bibinfo{year}{2017}), \bibinfo{pages}{268--278}.
\newblock
\showISSN{2373-9231}
\href{https://doi.org/10.1002/pra2.2017.14505401030}{doi:\nolinkurl{10.1002/pra2.2017.14505401030}}
\newblock
\shownote{\_eprint: https://onlinelibrary.wiley.com/doi/pdf/10.1002/pra2.2017.14505401030}.


\bibitem[McCarthy and Wright(2007)]%
        {mccarthy_technology_2007}
\bibfield{author}{\bibinfo{person}{John McCarthy} {and} \bibinfo{person}{Peter Wright}.} \bibinfo{year}{2007}\natexlab{}.
\newblock \bibinfo{booktitle}{\emph{Technology as experience} (\bibinfo{edition}{1. paperback ed} ed.)}.
\newblock \bibinfo{publisher}{MIT Press}, \bibinfo{address}{Cambridge, Mass}.
\newblock
\showISBNx{978-0-262-63355-0}


\bibitem[Meyer and Norman(2020)]%
        {meyer_changing_2020}
\bibfield{author}{\bibinfo{person}{Michael Meyer} {and} \bibinfo{person}{Don Norman}.} \bibinfo{year}{2020}\natexlab{}.
\newblock \showarticletitle{Changing {Design} {Education} for the 21st {Century}}.
\newblock \bibinfo{journal}{\emph{She Ji: The Journal of Design, Economics, and Innovation}} (\bibinfo{year}{2020}), \bibinfo{pages}{13--49}.
\newblock
\href{https://doi.org/10.1016/j.sheji.2019.12.002}{doi:\nolinkurl{10.1016/j.sheji.2019.12.002}}


\bibitem[Mikkelsen(2021)]%
        {mikkelsen_perceived_2021}
\bibfield{author}{\bibinfo{person}{Mogens~Frank Mikkelsen}.} \bibinfo{year}{2021}\natexlab{}.
\newblock \showarticletitle{Perceived project complexity: a survey among practitioners of project management}.
\newblock \bibinfo{journal}{\emph{International Journal of Managing Projects in Business}} \bibinfo{volume}{14}, \bibinfo{number}{3} (\bibinfo{date}{Jan.} \bibinfo{year}{2021}), \bibinfo{pages}{680--698}.
\newblock
\showISSN{1753-8378}
\href{https://doi.org/10.1108/IJMPB-03-2020-0095}{doi:\nolinkurl{10.1108/IJMPB-03-2020-0095}}
\newblock
\shownote{Publisher: Emerald Publishing Limited}.


\bibitem[Nelson and Stolterman(2002)]%
        {nelson_design_2002}
\bibfield{author}{\bibinfo{person}{H Nelson} {and} \bibinfo{person}{E Stolterman}.} \bibinfo{year}{2002}\natexlab{}.
\newblock \showarticletitle{Design judgment: decision making in the ‘real’ world}.
\newblock \bibinfo{journal}{\emph{Common Ground}} \bibinfo{volume}{6}, \bibinfo{number}{1} (\bibinfo{year}{2002}), \bibinfo{pages}{23--31}.
\newblock


\bibitem[Nelson and Stolterman(2012)]%
        {nelson_design_2012}
\bibfield{author}{\bibinfo{person}{Harold~G. Nelson} {and} \bibinfo{person}{Erik Stolterman}.} \bibinfo{year}{2012}\natexlab{}.
\newblock \bibinfo{booktitle}{\emph{The {Design} {Way}: {Intentional} {Change} in an {Unpredictable} {World}} (\bibinfo{edition}{2nd} ed.)}.
\newblock \bibinfo{publisher}{The MIT Press}, \bibinfo{address}{Cambridge, Massachusestts ; London, England}.
\newblock
\showISBNx{978-0-262-01817-3}


\bibitem[Norouzi et~al\mbox{.}(2015)]%
        {norouzi_architect_2015}
\bibfield{author}{\bibinfo{person}{Nima Norouzi}, \bibinfo{person}{Maryam Shabak}, \bibinfo{person}{Mohamed Rashid~Bin Embi}, {and} \bibinfo{person}{Tareef~Hayat Khan}.} \bibinfo{year}{2015}\natexlab{}.
\newblock \showarticletitle{The {Architect}, the {Client} and {Effective} {Communication} in {Architectural} {Design} {Practice}}.
\newblock \bibinfo{journal}{\emph{Procedia - Social and Behavioral Sciences}}  \bibinfo{volume}{172} (\bibinfo{date}{Jan.} \bibinfo{year}{2015}), \bibinfo{pages}{635--642}.
\newblock
\showISSN{18770428}
\href{https://doi.org/10.1016/j.sbspro.2015.01.413}{doi:\nolinkurl{10.1016/j.sbspro.2015.01.413}}


\bibitem[Ove~Hansson(1996)]%
        {ove_hansson_decision_1996}
\bibfield{author}{\bibinfo{person}{Sven Ove~Hansson}.} \bibinfo{year}{1996}\natexlab{}.
\newblock \showarticletitle{Decision {Making} {Under} {Great} {Uncertainty}}.
\newblock \bibinfo{journal}{\emph{Philosophy of the Social Sciences}} \bibinfo{volume}{26}, \bibinfo{number}{3} (\bibinfo{date}{Sept.} \bibinfo{year}{1996}), \bibinfo{pages}{369--386}.
\newblock
\showISSN{0048-3931, 1552-7441}
\href{https://doi.org/10.1177/004839319602600304}{doi:\nolinkurl{10.1177/004839319602600304}}


\bibitem[O’Riordan and Dahinden(2011)]%
        {oriordan_dealing_2011}
\bibfield{author}{\bibinfo{person}{Margaret O’Riordan} {and} \bibinfo{person}{Andre Dahinden}.} \bibinfo{year}{2011}\natexlab{}.
\newblock \showarticletitle{Dealing with uncertainty in general practice: an essential skill for the general practitioner}.
\newblock \bibinfo{journal}{\emph{Radcliffe Publishing}} (\bibinfo{year}{2011}).
\newblock


\bibitem[Parsons et~al\mbox{.}(2023)]%
        {parsons_developing_2023}
\bibfield{author}{\bibinfo{person}{Paul~C Parsons}, \bibinfo{person}{Prakash~Chandra Shukla}, \bibinfo{person}{Ali Baigelenov}, {and} \bibinfo{person}{Colin~M Gray}.} \bibinfo{year}{2023}\natexlab{}.
\newblock \showarticletitle{Developing {Framing} {Judgment} {Ability}: {Student} {Perceptions} from a {Graduate} {UX} {Design} {Program}}. In \bibinfo{booktitle}{\emph{Proceedings of the 5th {Annual} {Symposium} on {HCI} {Education}}}. \bibinfo{publisher}{ACM}, \bibinfo{address}{Hamburg Germany}, \bibinfo{pages}{23--32}.
\newblock
\showISBNx{9798400707377}
\href{https://doi.org/10.1145/3587399.3587401}{doi:\nolinkurl{10.1145/3587399.3587401}}


\bibitem[Pelz et~al\mbox{.}(2021)]%
        {pelz_mastering_2021}
\bibfield{editor}{\bibinfo{person}{Peter~F. Pelz}, \bibinfo{person}{Peter Groche}, \bibinfo{person}{Marc~E. Pfetsch}, {and} \bibinfo{person}{Maximilian Schaeffner}} (Eds.). \bibinfo{year}{2021}\natexlab{}.
\newblock \bibinfo{booktitle}{\emph{Mastering {Uncertainty} in {Mechanical} {Engineering}}}.
\newblock \bibinfo{publisher}{Springer Nature}.
\newblock
\href{https://doi.org/10.1007/978-3-030-78354-9}{doi:\nolinkurl{10.1007/978-3-030-78354-9}}
\newblock
\shownote{Accepted: 2021-10-13T13:52:56Z}.


\bibitem[Rittel(1967)]%
        {rittel_wicked_1967}
\bibfield{author}{\bibinfo{person}{Horst Rittel}.} \bibinfo{year}{1967}\natexlab{}.
\newblock \showarticletitle{Wicked {Problems}}. In \bibinfo{booktitle}{\emph{Management {Science}, ({December} 1967)}}, Vol.~\bibinfo{volume}{4(14)}.
\newblock


\bibitem[Rittel and Webber(1973)]%
        {rittel_dilemmas_1973}
\bibfield{author}{\bibinfo{person}{Horst Rittel} {and} \bibinfo{person}{Melvin Webber}.} \bibinfo{year}{1973}\natexlab{}.
\newblock \showarticletitle{Dilemmas in a {General} {Theory} of {Planning}}.
\newblock \bibinfo{journal}{\emph{Policy Sciences}} (\bibinfo{year}{1973}), \bibinfo{pages}{pp. 155--69}.
\newblock


\bibitem[Rocquigny et~al\mbox{.}(2008)]%
        {rocquigny_uncertainty_2008}
\bibfield{author}{\bibinfo{person}{Etienne~de Rocquigny}, \bibinfo{person}{Nicolas Devictor}, {and} \bibinfo{person}{Stefano Tarantola}.} \bibinfo{year}{2008}\natexlab{}.
\newblock \bibinfo{booktitle}{\emph{Uncertainty in {Industrial} {Practice}: {A} {Guide} to {Quantitative} {Uncertainty} {Management}}}.
\newblock \bibinfo{publisher}{John Wiley \& Sons}.
\newblock
\showISBNx{978-0-470-77074-0}
\newblock
\shownote{Google-Books-ID: qg8bqw6ByskC}.


\bibitem[Saunders et~al\mbox{.}(2015)]%
        {saunders_conceptualising_2015}
\bibfield{author}{\bibinfo{person}{Fiona~C. Saunders}, \bibinfo{person}{Andrew~W. Gale}, {and} \bibinfo{person}{Andrew~H. Sherry}.} \bibinfo{year}{2015}\natexlab{}.
\newblock \showarticletitle{Conceptualising uncertainty in safety-critical projects: {A} practitioner perspective}.
\newblock \bibinfo{journal}{\emph{International Journal of Project Management}} \bibinfo{volume}{33}, \bibinfo{number}{2} (\bibinfo{date}{Feb.} \bibinfo{year}{2015}), \bibinfo{pages}{467--478}.
\newblock
\showISSN{0263-7863}
\href{https://doi.org/10.1016/j.ijproman.2014.09.002}{doi:\nolinkurl{10.1016/j.ijproman.2014.09.002}}


\bibitem[Schön(1983)]%
        {schon_reflective_1983}
\bibfield{author}{\bibinfo{person}{Donald~A. Schön}.} \bibinfo{year}{1983}\natexlab{}.
\newblock \bibinfo{booktitle}{\emph{The reflective practitioner: how professionals think in action}}.
\newblock \bibinfo{publisher}{Basic Books}, \bibinfo{address}{New York}.
\newblock
\showISBNx{978-0-465-06878-4 978-0-465-06874-6}


\bibitem[Schön(1984)]%
        {schon_problems_1984}
\bibfield{author}{\bibinfo{person}{Donald~A. Schön}.} \bibinfo{year}{1984}\natexlab{}.
\newblock \showarticletitle{Problems, frames and perspectives on designing}.
\newblock \bibinfo{journal}{\emph{Design Studies}} \bibinfo{volume}{5}, \bibinfo{number}{3} (\bibinfo{year}{1984}), \bibinfo{pages}{132--136}.
\newblock
\showISSN{0142-694X}
\href{https://doi.org/10.1016/0142-694X(84)90002-4}{doi:\nolinkurl{10.1016/0142-694X(84)90002-4}}


\bibitem[Schön(1987)]%
        {schon_educating_1987}
\bibfield{author}{\bibinfo{person}{Donald~A. Schön}.} \bibinfo{year}{1987}\natexlab{}.
\newblock \showarticletitle{Educating the reflective practitioner}.
\newblock \bibinfo{journal}{\emph{Jossey-Bass San Francisco}} (\bibinfo{year}{1987}).
\newblock


\bibitem[Shukla et~al\mbox{.}(2025)]%
        {shukla_-skilling_2025}
\bibfield{author}{\bibinfo{person}{Prakash Shukla}, \bibinfo{person}{Phuong Bui}, \bibinfo{person}{Sean~S. Levy}, \bibinfo{person}{Max Kowalski}, \bibinfo{person}{Ali Baigelenov}, {and} \bibinfo{person}{Paul Parsons}.} \bibinfo{year}{2025}\natexlab{}.
\newblock \bibinfo{title}{De-skilling, {Cognitive} {Offloading}, and {Misplaced} {Responsibilities}: {Potential} {Ironies} of {AI}-{Assisted} {Design}}.
\newblock
\href{https://doi.org/10.1145/3706599.3719931}{doi:\nolinkurl{10.1145/3706599.3719931}}
\newblock
\shownote{arXiv:2503.03924 [cs]}.


\bibitem[Shukla et~al\mbox{.}(2024)]%
        {shukla_communication_2024}
\bibfield{author}{\bibinfo{person}{Prakash Shukla}, \bibinfo{person}{Suchismita Naik}, \bibinfo{person}{Ike Obi}, \bibinfo{person}{Phuong Bui}, {and} \bibinfo{person}{Paul Parsons}.} \bibinfo{year}{2024}\natexlab{}.
\newblock \showarticletitle{Communication {Challenges} {Reported} by {UX} {Designers} on {Social} {Media}: {An} {Analysis} of {Subreddit} {Discussions}}. In \bibinfo{booktitle}{\emph{Extended {Abstracts} of the 2024 {CHI} {Conference} on {Human} {Factors} in {Computing} {Systems}}} \emph{(\bibinfo{series}{{CHI} {EA} '24})}. \bibinfo{publisher}{Association for Computing Machinery}, \bibinfo{address}{New York, NY, USA}, \bibinfo{pages}{1--6}.
\newblock
\showISBNx{9798400703317}
\href{https://doi.org/10.1145/3613905.3650881}{doi:\nolinkurl{10.1145/3613905.3650881}}


\bibitem[Simon(1973)]%
        {simon_structure_1973}
\bibfield{author}{\bibinfo{person}{Herbert~A. Simon}.} \bibinfo{year}{1973}\natexlab{}.
\newblock \showarticletitle{The structure of ill structured problems}.
\newblock \bibinfo{journal}{\emph{Artificial intelligence}} \bibinfo{volume}{4}, \bibinfo{number}{3-4} (\bibinfo{year}{1973}), \bibinfo{pages}{181--201}.
\newblock
\urldef\tempurl%
\url{https://www.sciencedirect.com/science/article/pii/0004370273900118}
\showURL{%
\tempurl}
\newblock
\shownote{Publisher: Elsevier}.


\bibitem[Sniazhko(2019)]%
        {sniazhko_uncertainty_2019}
\bibfield{author}{\bibinfo{person}{Sniazhana Sniazhko}.} \bibinfo{year}{2019}\natexlab{}.
\newblock \showarticletitle{Uncertainty in decision-making: {A} review of the international business literature}.
\newblock \bibinfo{journal}{\emph{Cogent Business \& Management}} \bibinfo{volume}{6}, \bibinfo{number}{1} (\bibinfo{date}{Jan.} \bibinfo{year}{2019}), \bibinfo{pages}{1650692}.
\newblock
\showISSN{2331-1975}
\href{https://doi.org/10.1080/23311975.2019.1650692}{doi:\nolinkurl{10.1080/23311975.2019.1650692}}


\bibitem[Soden et~al\mbox{.}(2022)]%
        {soden_modes_2022}
\bibfield{author}{\bibinfo{person}{Robert Soden}, \bibinfo{person}{Laura Devendorf}, \bibinfo{person}{Richmond Wong}, \bibinfo{person}{Yoko Akama}, {and} \bibinfo{person}{Ann Light}.} \bibinfo{year}{2022}\natexlab{}.
\newblock \showarticletitle{Modes of {Uncertainty} in {HCI}}.
\newblock \bibinfo{journal}{\emph{Foundations and Trends® in Human–Computer Interaction}} \bibinfo{volume}{15}, \bibinfo{number}{4} (\bibinfo{date}{Aug.} \bibinfo{year}{2022}), \bibinfo{pages}{317--426}.
\newblock
\showISSN{1551-3955, 1551-3963}
\href{https://doi.org/10.1561/1100000085}{doi:\nolinkurl{10.1561/1100000085}}
\newblock
\shownote{Publisher: Now Publishers, Inc.}.


\bibitem[Stolterman(2008)]%
        {stolterman_nature_2008}
\bibfield{author}{\bibinfo{person}{Erik Stolterman}.} \bibinfo{year}{2008}\natexlab{}.
\newblock \showarticletitle{The {Nature} of {Design} {Practice} and {Implications} for {Interaction} {Design} {Research}}.
\newblock \bibinfo{journal}{\emph{International Journal of Design}} \bibinfo{volume}{2}, \bibinfo{number}{1} (\bibinfo{year}{2008}), \bibinfo{pages}{55--65}.
\newblock
\showISSN{09501991}
\href{https://doi.org/10.1016/j.phymed.2007.09.005}{doi:\nolinkurl{10.1016/j.phymed.2007.09.005}}


\bibitem[Stolterman(2021)]%
        {stolterman_challenge_2021}
\bibfield{author}{\bibinfo{person}{Erik Stolterman}.} \bibinfo{year}{2021}\natexlab{}.
\newblock \showarticletitle{The {Challenge} of {Improving} {Designing}}.
\newblock  \bibinfo{volume}{15}, \bibinfo{number}{1} (\bibinfo{year}{2021}).
\newblock


\bibitem[Suchman(1987)]%
        {suchman_plans_1987}
\bibfield{author}{\bibinfo{person}{Lucille~Alice Suchman}.} \bibinfo{year}{1987}\natexlab{}.
\newblock \bibinfo{booktitle}{\emph{Plans and {Situated} {Actions}: {The} {Problem} of {Human}-{Machine} {Communication}}}.
\newblock \bibinfo{publisher}{Cambridge University Press}.
\newblock
\showISBNx{978-0-521-33739-7}
\newblock
\shownote{Google-Books-ID: AJ\_eBJtHxmsC}.


\bibitem[Swain(2018)]%
        {swain_hybrid_2018}
\bibfield{author}{\bibinfo{person}{Jon Swain}.} \bibinfo{year}{2018}\natexlab{}.
\newblock \showarticletitle{A hybrid approach to thematic analysis in qualitative research: {Using} a practical example}.
\newblock In \bibinfo{booktitle}{\emph{Sage {Research} {Methods}}}. \bibinfo{series}{Cases}, Vol.~\bibinfo{volume}{Part 2}. \bibinfo{publisher}{SAGE Publications, Ltd.}
\newblock


\bibitem[Sánchez~Chamorro et~al\mbox{.}(2023)]%
        {sanchez_chamorro_ethical_2023}
\bibfield{author}{\bibinfo{person}{Lorena Sánchez~Chamorro}, \bibinfo{person}{Kerstin Bongard-Blanchy}, {and} \bibinfo{person}{Vincent Koenig}.} \bibinfo{year}{2023}\natexlab{}.
\newblock \showarticletitle{Ethical {Tensions} in {UX} {Design} {Practice}: {Exploring} the {Fine} {Line} {Between} {Persuasion} and {Manipulation} in {Online} {Interfaces}}. In \bibinfo{booktitle}{\emph{Proceedings of the 2023 {ACM} {Designing} {Interactive} {Systems} {Conference}}}. \bibinfo{publisher}{ACM}, \bibinfo{address}{Pittsburgh PA USA}, \bibinfo{pages}{2408--2422}.
\newblock
\showISBNx{978-1-4503-9893-0}
\href{https://doi.org/10.1145/3563657.3596013}{doi:\nolinkurl{10.1145/3563657.3596013}}


\bibitem[Takaffoli et~al\mbox{.}(2024)]%
        {takaffoli_generative_2024}
\bibfield{author}{\bibinfo{person}{Macy Takaffoli}, \bibinfo{person}{Sijia Li}, {and} \bibinfo{person}{Ville Mäkelä}.} \bibinfo{year}{2024}\natexlab{}.
\newblock \showarticletitle{Generative {AI} in {User} {Experience} {Design} and {Research}: {How} {Do} {UX} {Practitioners}, {Teams}, and {Companies} {Use} {GenAI} in {Industry}?}. In \bibinfo{booktitle}{\emph{Proceedings of the 2024 {ACM} {Designing} {Interactive} {Systems} {Conference}}} \emph{(\bibinfo{series}{{DIS} '24})}. \bibinfo{publisher}{Association for Computing Machinery}, \bibinfo{address}{New York, NY, USA}, \bibinfo{pages}{1579--1593}.
\newblock
\showISBNx{979-8-4007-0583-0}
\href{https://doi.org/10.1145/3643834.3660720}{doi:\nolinkurl{10.1145/3643834.3660720}}
\newblock
\shownote{event-place: Copenhagen, Denmark}.


\bibitem[Tracey and Hutchinson(2016)]%
        {tracey_uncertainty_2016}
\bibfield{author}{\bibinfo{person}{Monica~W. Tracey} {and} \bibinfo{person}{Alisa Hutchinson}.} \bibinfo{year}{2016}\natexlab{}.
\newblock \showarticletitle{Uncertainty, reflection, and designer identity development}.
\newblock \bibinfo{journal}{\emph{Design Studies}}  \bibinfo{volume}{42} (\bibinfo{date}{Jan.} \bibinfo{year}{2016}), \bibinfo{pages}{86--109}.
\newblock
\showISSN{0142-694X}
\href{https://doi.org/10.1016/j.destud.2015.10.004}{doi:\nolinkurl{10.1016/j.destud.2015.10.004}}


\bibitem[Vinney(2022)]%
        {vinney_11_2022}
\bibfield{author}{\bibinfo{person}{Cynthia Vinney}.} \bibinfo{year}{2022}\natexlab{}.
\newblock \bibinfo{title}{11 popular {UX} job roles and titles explained - {UX} {Design} {Institute}}.
\newblock
\urldef\tempurl%
\url{https://www.uxdesigninstitute.com/blog/ux-job-roles/}
\showURL{%
\tempurl}
\newblock
\shownote{Section: Breaking into UX}.


\bibitem[Vorvoreanu et~al\mbox{.}(2017)]%
        {vorvoreanu_advancing_2017}
\bibfield{author}{\bibinfo{person}{Mihaela Vorvoreanu}, \bibinfo{person}{Colin~M. Gray}, \bibinfo{person}{Paul Parsons}, {and} \bibinfo{person}{Nancy Rasche}.} \bibinfo{year}{2017}\natexlab{}.
\newblock \showarticletitle{Advancing {UX} {Education}: {A} {Model} for {Integrated} {Studio} {Pedagogy}}. In \bibinfo{booktitle}{\emph{Proceedings of the 2017 {CHI} {Conference} on {Human} {Factors} in {Computing} {Systems}}} \emph{(\bibinfo{series}{{CHI} '17})}. \bibinfo{publisher}{ACM}, \bibinfo{address}{New York, NY, USA}, \bibinfo{pages}{1441--1446}.
\newblock
\showISBNx{978-1-4503-4655-9}
\href{https://doi.org/10.1145/3025453.3025726}{doi:\nolinkurl{10.1145/3025453.3025726}}


\bibitem[Walker and Marchau(2003)]%
        {walker_dealing_2003}
\bibfield{author}{\bibinfo{person}{W.~E. Walker} {and} \bibinfo{person}{V.~A. W.~J. Marchau}.} \bibinfo{year}{2003}\natexlab{}.
\newblock \showarticletitle{Dealing {With} {Uncertainty} in {Policy} {Analysis} and {Policymaking}}.
\newblock \bibinfo{journal}{\emph{Integrated Assessment}} (\bibinfo{date}{March} \bibinfo{year}{2003}).
\newblock
\href{https://doi.org/10.1076/iaij.4.1.1.16462}{doi:\nolinkurl{10.1076/iaij.4.1.1.16462}}
\newblock
\shownote{Publisher: Taylor \& Francis Group}.


\bibitem[Watkins et~al\mbox{.}(2020)]%
        {watkins_tensions_2020}
\bibfield{author}{\bibinfo{person}{Christopher~Rhys Watkins}, \bibinfo{person}{Colin~M. Gray}, \bibinfo{person}{Austin~L. Toombs}, {and} \bibinfo{person}{Paul Parsons}.} \bibinfo{year}{2020}\natexlab{}.
\newblock \showarticletitle{Tensions in {Enacting} a {Design} {Philosophy} in {UX} {Practice}}. In \bibinfo{booktitle}{\emph{Proceedings of the 2020 {ACM} {Designing} {Interactive} {Systems} {Conference}}}. \bibinfo{publisher}{ACM}, \bibinfo{address}{Eindhoven Netherlands}, \bibinfo{pages}{2107--2118}.
\newblock
\showISBNx{978-1-4503-6974-9}
\href{https://doi.org/10.1145/3357236.3395505}{doi:\nolinkurl{10.1145/3357236.3395505}}


\bibitem[Woods(2006)]%
        {woods_essential_2006}
\bibfield{author}{\bibinfo{person}{David~D. Woods}.} \bibinfo{year}{2006}\natexlab{}.
\newblock \showarticletitle{Essential {Characteristics} of {Resilience}}.
\newblock In \bibinfo{booktitle}{\emph{Resilience {Engineering}}}. \bibinfo{publisher}{CRC Press}.
\newblock
\showISBNx{978-1-315-60568-5}
\newblock
\shownote{Num Pages: 14}.


\bibitem[Yang et~al\mbox{.}(2018)]%
        {yang_investigating_2018}
\bibfield{author}{\bibinfo{person}{Qian Yang}, \bibinfo{person}{Alex Scuito}, \bibinfo{person}{John Zimmerman}, \bibinfo{person}{Jodi Forlizzi}, {and} \bibinfo{person}{Aaron Steinfeld}.} \bibinfo{year}{2018}\natexlab{}.
\newblock \showarticletitle{Investigating {How} {Experienced} {UX} {Designers} {Effectively} {Work} with {Machine} {Learning}}. In \bibinfo{booktitle}{\emph{Proceedings of the 2018 {Designing} {Interactive} {Systems} {Conference}}} \emph{(\bibinfo{series}{{DIS} '18})}. \bibinfo{publisher}{Association for Computing Machinery}, \bibinfo{address}{New York, NY, USA}, \bibinfo{pages}{585--596}.
\newblock
\showISBNx{978-1-4503-5198-0}
\href{https://doi.org/10.1145/3196709.3196730}{doi:\nolinkurl{10.1145/3196709.3196730}}


\bibitem[Zdanowska and Taylor(2022)]%
        {zdanowska_study_2022}
\bibfield{author}{\bibinfo{person}{Sabah Zdanowska} {and} \bibinfo{person}{Alex~S Taylor}.} \bibinfo{year}{2022}\natexlab{}.
\newblock \showarticletitle{A study of {UX} practitioners roles in designing real-world, enterprise {ML} systems}. In \bibinfo{booktitle}{\emph{{CHI} {Conference} on {Human} {Factors} in {Computing} {Systems}}}. \bibinfo{publisher}{ACM}, \bibinfo{address}{New Orleans LA USA}, \bibinfo{pages}{1--15}.
\newblock
\showISBNx{978-1-4503-9157-3}
\href{https://doi.org/10.1145/3491102.3517607}{doi:\nolinkurl{10.1145/3491102.3517607}}


\end{thebibliography}


\end{document}